\documentclass[prl,twocolumn,floatix,amssymb,showpacs,amsmath, superscriptaddress,longbibliography]{revtex4-2}
\usepackage{graphicx}
\usepackage{epsfig}
\usepackage{amsmath,amssymb,amsthm}
\usepackage[utf8]{inputenc}
\usepackage[dvipsnames]{xcolor}
\usepackage[colorlinks, urlcolor=blue, linkcolor=red]{hyperref}

\usepackage{soul}

\newcommand{\beqa}{\begin{eqnarray}}
\newcommand{\eeqa}{\end{eqnarray}}

\newcommand{\crea}[1]{{\hat #1}^\dagger}
\newcommand{\ann}[1]{\hat #1}
\newcommand{\bra}[1]{\langle #1\rvert}
\newcommand{\ket}[1]{\lvert#1\rangle}

\newcommand{\bracket}[2]{\langle#1|#2\rangle}

\newcommand{\expec}[1]{\left\langle #1 \right\rangle}
\newcommand{\LL}{\mathcal{L}}
\newcommand{\DD}{\mathcal{D}}
\newcommand{\rhot}{\hat{\rho}(t)}
\newcommand{\de}{{\rm d}}

\def\aop{\hat{a}}
\def\bop{\hat{b}}
\def\cop{\hat{c}}
\def\adag{\hat{a}^\dagger}
\def\bdag{\hat{b}^\dagger}
\def\cdag{\hat{c}^\dagger}
\def\xop{\hat{x}}
\def\yop{\hat{y}}
\def\zop{\hat{z}}
\def\pop{\hat{p}}
\def\uop{\hat{u}}
\def\Hop{\hat{H}}

\renewcommand{\boxed}[2]{\textcolor{#1}{%
\tikz[baseline={([yshift=-1ex]current bounding box.center)}] \node [rectangle, minimum width=1ex,rounded corners,draw] {\normalcolor\m@th$\displaystyle#2$};}}

\newcounter{appsection}
\newcounter{appsubsection}[appsection]

\usepackage{bookmark}
\usepackage[normalem]{ulem}

\newcommand\redsout{\bgroup\markoverwith{\textcolor{red}{\rule[0.5ex]{2pt}{2pt}}}\ULon}
\newcommand\bluesout{\bgroup\markoverwith{\textcolor{blue}{\rule[0.5ex]{2pt}{2pt}}}\ULon}
\newcommand\greensout{\bgroup\markoverwith{\textcolor{green}{\rule[0.5ex]{2pt}{2pt}}}\ULon}

\begin{document}

\title{Non-Gaussian superradiant transition via three-body ultrastrong coupling}
\author{Fabrizio Minganti}
\affiliation{Institute of Physics, Ecole Polytechnique F\'ed\'erale de Lausanne (EPFL), CH-1015 Lausanne, Switzerland}
\affiliation{Center for Quantum Science and Engineering, Ecole Polytechnique F\'ed\'erale de Lausanne (EPFL), CH-1015 Lausanne, Switzerland}
 
\author{Louis Garbe}
\affiliation{Vienna Center for Quantum Science and Technology, Atominstitut, TU Wien, 1040 Vienna, Austria}

\author{Alexandre Le Boit\'e} 
\email{alexande.leboite@univ-paris-diderot.fr}
\affiliation{Universit\'{e} Paris Cit\'{e}, Laboratoire Mat\'{e}riaux et Ph\'{e}nom\`{e}nes Quantiques (MPQ), CNRS-UMR7162, F-75013 Paris, France}

\author{Simone Felicetti} 
\affiliation{Istituto di Fotonica e Nanotecnologie, Consiglio Nazionale delle Ricerche (IFN-CNR), 00156 Roma, Italy}

\begin{abstract}
We introduce a class of quantum optical Hamiltonian characterized by three-body couplings, and propose a circuit-QED scheme based on state-of-the-art technology that implements the considered model. Unlike two-body light-matter interactions, this three-body coupling Hamiltonian is exclusively composed of terms which do not conserve the particle number.
We explore the three-body ultrastrong coupling regime, showing
the emergence of a superradiant phase transition which is of first order, is characterized by the breaking of a $\mathbb{Z}_2\times \mathbb{Z}_2$ symmetry, and has a strongly non-Gaussian nature.
Indeed, in contrast to what is observed in any two-body-coupling model, in proximity of the transition the ground state exhibits a divergent coskewness, i.e., quantum correlations that cannot be captured within semiclassical  and Gaussian approximations. Furthermore, we demonstrate the robustness of our findings by including dissipative processes in the model, showing that the steady-state of the system inherits from the ground states the most prominent features of the transition.
\end{abstract}

\maketitle

\paragraph{ Introduction.---} Controlling the interaction between light and matter is one of the main research axes of modern quantum science. It has far-reaching implications for fundamental research and practical applications in quantum optics~\cite{HarocheLecture}, condensed matter~\cite{paravicini2019magneto,cortese2021excitons}, and polaritonic chemistry~\cite{herrera2020molecular,fregoni2021theoretical}. The achievement of the strong coupling regime, where the interaction strength overcomes losses, 
led to the observation of quantum-coherent energy exchanges and paved the way to a plethora of applications in quantum technologies. When the coupling strength is further increased, becoming comparable to the bare-system frequencies, the ultrastrong coupling (USC) regime is reached~\cite{Forn-Diaz2019,Frisk-Kockum2019,LeBoite2020}, leading to deep modifications of optical, material, and chemical properties.

One of the most debated theoretical predictions regarding the USC regime is the emergence of a \textit{superradiant} phase transition driven by quantum light-matter interaction~\cite{Kirton2019}. When increasing the coupling strength, the ground state transitions from the vacuum to a superradiant phase populated by a macroscopic number of photonic excitations. 
Despite these theoretical predictions, the presence of renormalizing terms in realistic physical settings arguably prevents the emergence of the superradiant phase at equilibrium~\cite{Bernardis18,stokes2019gauge,distefano2019resolution,Andolina19,nataf2010no,garcia2015light,Manucharyan17,DeBernardis18_2}. However, this issue can be circumvented using optical pumping schemes and analog quantum simulation techniques~\cite{Georgescu2014}, where the light-matter coupling is effectively enhanced and pushed into the USC regime. 
In the last few years, this approach has been successfully implemented in circuit QED~\cite{Langford2017, Braumuller2017,Markovic2018}, in trapped ions~\cite{Lv2018}, opto- and electro-mechanical devices~\cite{Peterson2019} and atomic systems~\cite{Dareau2018,mivehvar2021cavity}, where it led to the observation of superradiant transitions~\cite{Black03,Baumann2010,Zhiqiang2017} using driven ultracold atoms in an optical cavity.

These effective implementations of USC can reach extreme regimes of parameters, and phase transitions can emerge in systems with a finite number of components \cite{AshhabPRA13,Hwang2015,Liou17,Hwang2018}, where the thermodynamic limit is substituted by a rescaling of the parameters. These finite-component phase transitions are easier to control~\cite{Puebla2017} than their many-body counterparts and offer an interesting framework for the study of critical phenomena~\cite{Larson2017,peng_unified2019,Innocenti_2020,ying2022quantum,zhao2021frustrated,PalacinoPRR21,BaksicPRL14}. For instance, it was recently shown~\cite{Felicetti2020} that the features of a superradiant phase are universally determined by key spectral properties of the model, and thus by the underlying symmetry of the USC interaction.
Beyond their fundamental interest, finite-component phase transitions open perspectives for quantum technologies. Notably, finite-component critical phenomena in atomic and solid-state devices are promising candidates for the development of critical quantum sensors~\cite{Garbe2020,Ivanov2020,Wald2020,Gietka2021adiabaticcritical,Chu2021,DiCandia2021critical,Salado2021,garbe2021critical,gietka2021exponentially,garbe2021exponential,gietka2021squeezing,ilias2021criticality}.

Here, we introduce a novel kind of superradiant phase transition, induced by three-body coupling in the USC regime, and design a scheme to implement this phenomenology based on recent experimental developments \cite{Chang2020}.
Our model consists of three nonlinear quantum resonators, whose coupling Hamiltonian has a $\mathbb{Z}_2\times \mathbb{Z}_2$ symmetry. 
We use both analytical and numerical tools to characterize the unconventional properties of its finite-component phase transition. With respect to standard superradiant transitions induced by two-body interactions, the most prominent features of our model are: (i) The higher-order degeneracy of the ground state in the superradiant phase  (ii) The non-Gaussian nature of the ground state at the transition, as certified by the coskewness of the photon statistics. We then present a microwave-pumping scheme that implements the considered model in a circuit-QED device, feasible with current technology. Finally, we provide analytical and numerical evidence that the key features of the phase transition are preserved in the presence of drive and dissipation, showing that this novel phenomenology is of direct experimental relevance.

\begin{figure}
    \centering
    \includegraphics[ width=0.4\linewidth]{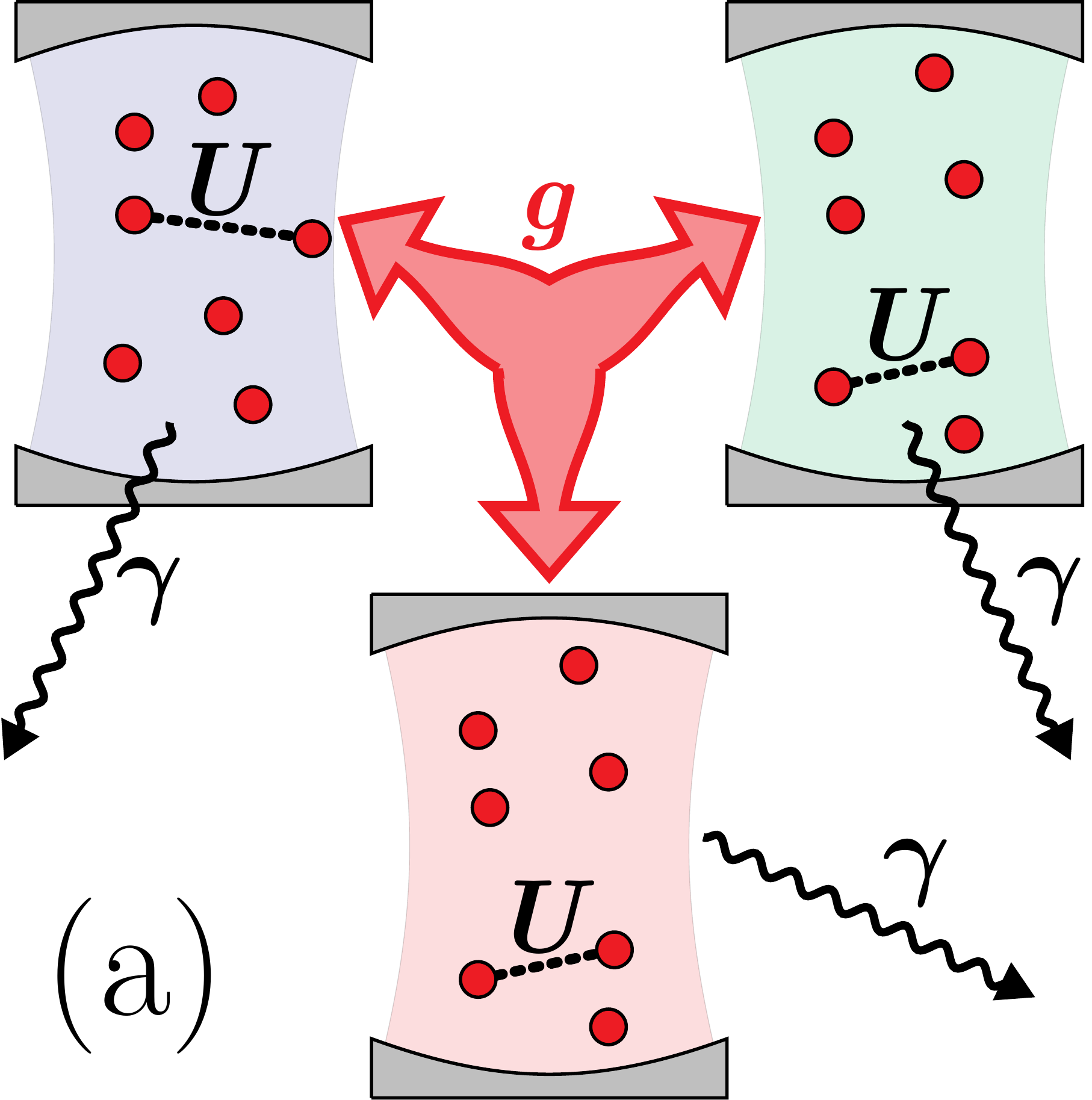}
    \includegraphics[width=0.58\linewidth]{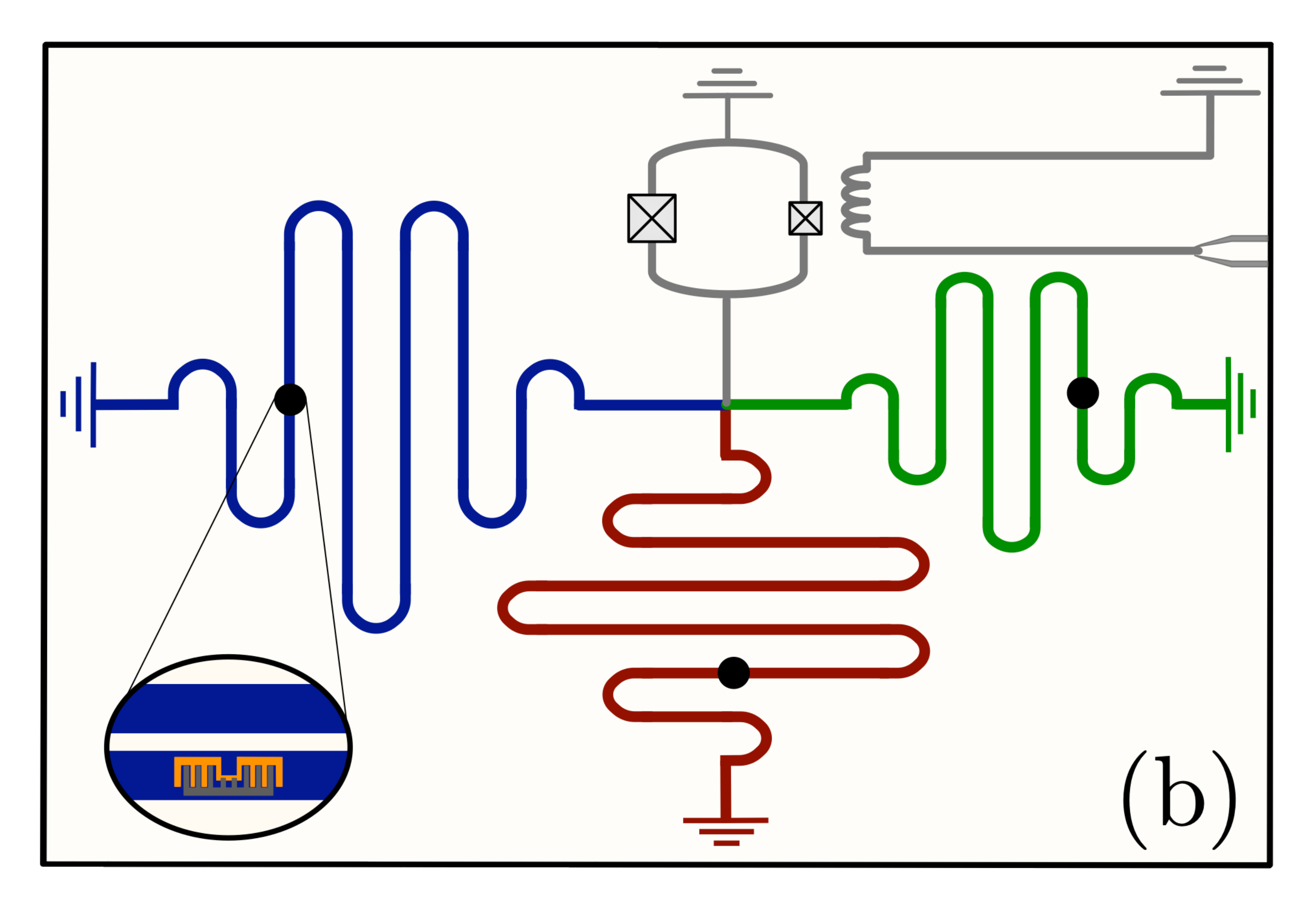}    
    \includegraphics[width=0.49\textwidth]{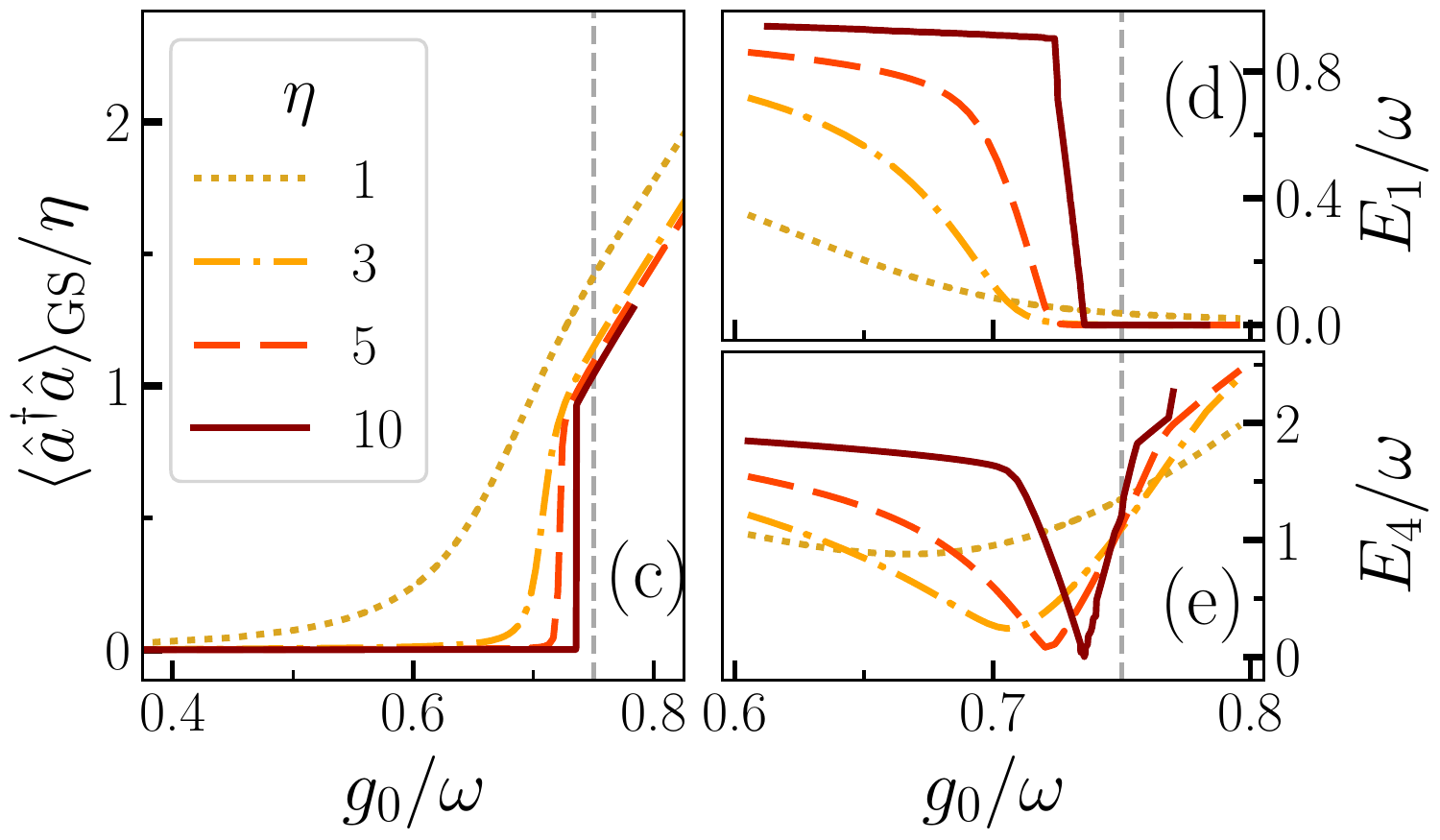}
    \caption{(a) Schematic representation of a system characterized by three-body USC realized, e.g., by the superconducting circuit in (b). (c) Rescaled photon number $\expec{\hat{a}^\dagger \hat{a}}/\eta$ in the ground state [c.f. Eq.~\eqref{Eq:Scaling}] vs the three-body coupling strength $g_0$. (d,e) Energy of the first excited state $E_1$ -- coinciding with $E_2$ and $E_3$ -- and that of the fourth excited state $E_4$. While $E_1$ captures the spontaneous symmetry breaking, $E_4$ is associated with the first-order discontinuity.
    The first-order transition point predicted by the semiclassical theory is indicated by a vertical dashed line in (c-e).
    We set $U_0/\omega=1$.
    }
    \label{fig:fig_1}
\end{figure}

\paragraph{Three-body coupling Hamiltonian.---} We consider a trimer of non-linear resonators coupled via a three-body coupling term. Among the different possible designs offered by quantum simulation techniques [see Figs.~\ref{fig:fig_1}(a,b)], we focus in this work on the following Hamiltonian 
\begin{equation}
\label{Eq:Trimer_Hamiltonian}
\begin{split}
    \hat{H} &= \omega \crea{a}\ann{a} + \omega \crea{b}\ann{b} + \omega \crea{c}\ann{c}  + U \crea{a}\crea{a}\ann{a}\ann{a} +U\crea{b}\crea{b}\ann{b}\ann{b}  \\ &+U\crea{c}\crea{c}\ann{c}\ann{c} + g \left(\crea{a} + \ann{a} \right) \left(\crea{b} + \ann{b} \right) \left(\crea{c} + \ann{c} \right),
\end{split}
\end{equation}
which is a three-body generalization of standard dipolar couplings.
We defined the annihilation operators $\ann{a}$, $\ann{b}$, and $\ann{c}$ of three bosonic modes with bare frequencies $\omega$, while $U$ is the on-site nonlinearity and $g$ is the interaction strength. Notice that none of the coupling terms in Eq.~\eqref{Eq:Trimer_Hamiltonian} induces resonant transitions, i.e., \textit{all terms are fast-oscillating in the interaction picture}. However, for three-body interactions the onset of the USC regime -- where counter-rotating terms become relevant -- takes place for very low values of $g$. 
We  show in what follows how the superradiant states emerging in the USC regime are strongly constrained by symmetry properties imposed by the specific form of the interaction terms.  
For the considered model, the Hamiltonian commutes with the following operators
$\hat{S}_1= e^{i \pi (\crea{a}\ann{a}+ \crea{b}\ann{b})}, \; \hat{S}_2= e^{i \pi (\crea{a}\ann{a}+ \crea{c}\ann{c})}, \; \hat{S}_3= e^{i \pi (\crea{b}\ann{b}+ \crea{c}\ann{c})}$.
Since $\hat{S}_3=\hat{S}_1 \hat{S}_2$ and $[\hat{S}_1, \hat{S}_2]=0$, the eigenstates $\ket{\Psi_i}$ are characterized by a $\mathbb{Z}_2\otimes \mathbb{Z}_2$ symmetry, i.e. two quantum numbers $(s_1, s_2)=(\pm 1, \pm 1)$ such that $\hat{S}_{1, \, 2} \ket{\Psi_i}= s_{1, \, 2} \ket{\Psi_i}$.

\paragraph{Superradiant transition.---}~
Generally speaking, when the interaction strength $g$ is increased up to values where the Dicke-coupling term is dominant ($g\gg \omega,U$), a superradiant state always emerges after a cross-over~\cite{Felicetti2020}. To convert the crossover into a phase transition, an effective thermodynamic limit must be introduced~\cite{Hwang2015,peng_unified2019}. We identify here a parameter-scaling limit that induces a critical transition from the vacuum to a superradiant phase in the considered finite-component setup. We introduce the effective parameter $\eta$ and the following scaling laws,
\begin{align}\label{Eq:Scaling}
g = \frac{g_0}{\sqrt{\eta}},\quad U  = \frac{U_0}{\eta},
\end{align}
the thermodynamic limit being $\eta\to\infty$.
This choice ensures that all terms in the Hamiltonian scale similarly with $\eta$ in the superradiant phase, where we expect $\expec{\ann{a}},\langle\ann{b}\rangle,\expec{\ann{c}} \sim \sqrt{\eta}$. 
Let us first investigate the transition at the semiclassical level. We search for energy minima of the mean-field potential resulting from the substitution $\ann{a} \to \alpha$, $\ann{b} \to \beta$, $\ann{c} \to \gamma$ in Eq.~\eqref{Eq:Trimer_Hamiltonian}, where $\alpha,\beta, \gamma \in \mathbb{C}$. 
Under this approximation the normal-to-superradiant phase transition is driven by the $\eta$-independent parameter $\lambda = g \sqrt{2/(\omega U)}$. 
We identify three regimes:
i) For  $\lambda < 1$,  the system is the normal phase (vacuum). There is no superradiant extremum in $ H(\alpha,\beta,\gamma)$.
ii) For $1<\lambda <3/({2\sqrt{2}})$,  $H(\alpha,\beta,\gamma)$ has four superradiant local minima. These states are directly related to the $\mathbb{Z}_2\times \mathbb{Z}_2$ symmetry. They are obtained by applying the symmetry operators $\hat{S}_1$, $\hat{S}_2$, and $\hat{S}_3$ on a coherent state of the form $\ket{\bar{X},\bar{X},\bar{X}}$, where $\bar{X} = -\sqrt{\frac{\omega \eta}{2\epsilon_0}}(\lambda+\sqrt{\lambda^2-1})$. However, the global minimum is still the vacuum.
iii) For $ \lambda>  {3}/({2\sqrt{2}})  $, the superradiant minima become energetically favourable with respect to the vacuum state, and a  first-order transition to a superradiant 4-fold degenerate ground state occurs. At this level of analysis we already see that both the order of the transition and the degeneracy of the ground state are modified with respect to the $\mathbb{Z}_2$ Dicke-like superradiant transitions.  Note that the expected scaling $
\expec{\ann{a}},\langle\ann{b}\rangle,\expec{\ann{c}} \propto \sqrt{\eta}$
is recovered in the superradiant phase.

We now extend the analytical description including quantum fluctuations via a standard Bogoliubov approach \cite{Carusotto2013}. Namely, we expand the Hamiltonian around the mean-field solutions according to 
$\ann{a} \to \bar{X}+\hat{\mu} $, $\ann{b} \to \bar{X} + \hat{\nu}$ and $
\ann{c} \to \bar{X}+\hat{\zeta}$.
Keeping only second-order terms in $\hat{\mu}$,$\hat{\nu}$, and $\hat{\zeta}$ leads to a quadratic Hamiltonian which can be readily diagonalized~\cite{SUPMAT}. %
We find that there is a region $ \lambda \in ]1,1+l]$,  where $l = \left(\frac{U_0 }{\eta\omega}\right)^{4/5} $, in which fluctuations of the superradiant states are relevant. 
Thus, the semiclassical picture can be completed as:
i) When $1<\lambda <1+l $, quantum fluctuations make the superradiant states unstable.
ii) For $1+ l<\lambda <{3}/({2\sqrt{2}})$, superradiant local minima are not yet the ground state, but are stable. 
iii) For $\lambda >{3}/({2\sqrt{2}})$, the superradiant states become global minima, and the phase transition takes place. 
This treatment predicts that the fluctuations in both the normal and superradiant states are bounded at the phase transition. This is an important difference with respect to  Dicke-like phase transitions, where similar Gaussian treatments predict that the transition is accompanied by large quantum fluctuations. %

As presented in Fig.~\ref{fig:fig_1}, the mean-field theory correctly predicts some features of the phase transition, as confirmed by numerical simulations \cite{SUPMAT}. In Fig.~\ref{fig:fig_1} (c) we show the mean photon number in the ground state as a function of the coupling strength $g$ for different values of $\eta$. Even for relatively small $\eta$, the results give a clear signature of a first-order phase transition occurring at  $\lambda = {3}/({2\sqrt{2}})$ (i.e., $g_0/\omega = 0.75$ in Fig.~\ref{fig:fig_1}). 
The first-order discontinuity is also revealed by the behavior of the energy of the fourth excited state ($\hat{H}\ket{\Psi_4} = E_4 \ket{\Psi_4}$) plotted in Fig.~\ref{fig:fig_1} (e). The state $\ket{\Psi_4}$ has the same symmetry properties as $\ket{\Psi_{\rm GS}}$ because $\hat{S}_{1, \,2} \ket{\Psi_4} = \ket{\Psi_4}$, and therefore the avoided level crossing shown in Fig.~\ref{fig:fig_1} (e) is the precursor of the true criticality~\cite{BinderPRB84} emerging in the thermodynamic limit $\eta\to\infty$.
The breaking of the $\mathbb{Z}_2\times \mathbb{Z}_2$ symmetry is evidenced by the change in the energy spectrum at the transition of $\ket{\Psi_{1, \, 2, \, 3}}$, i.e., the eigenvectors belonging to different symmetry sectors with respect to the ground state. As shown in Fig.~\ref{fig:fig_1} (d), the energy gap between the ground state and the first excited states closes and the ground state becomes almost four-fold degenerate for $\lambda \geq  {3}/({2\sqrt{2}})$. 

\begin{figure}
    \centering
    \includegraphics[width=0.49\textwidth]{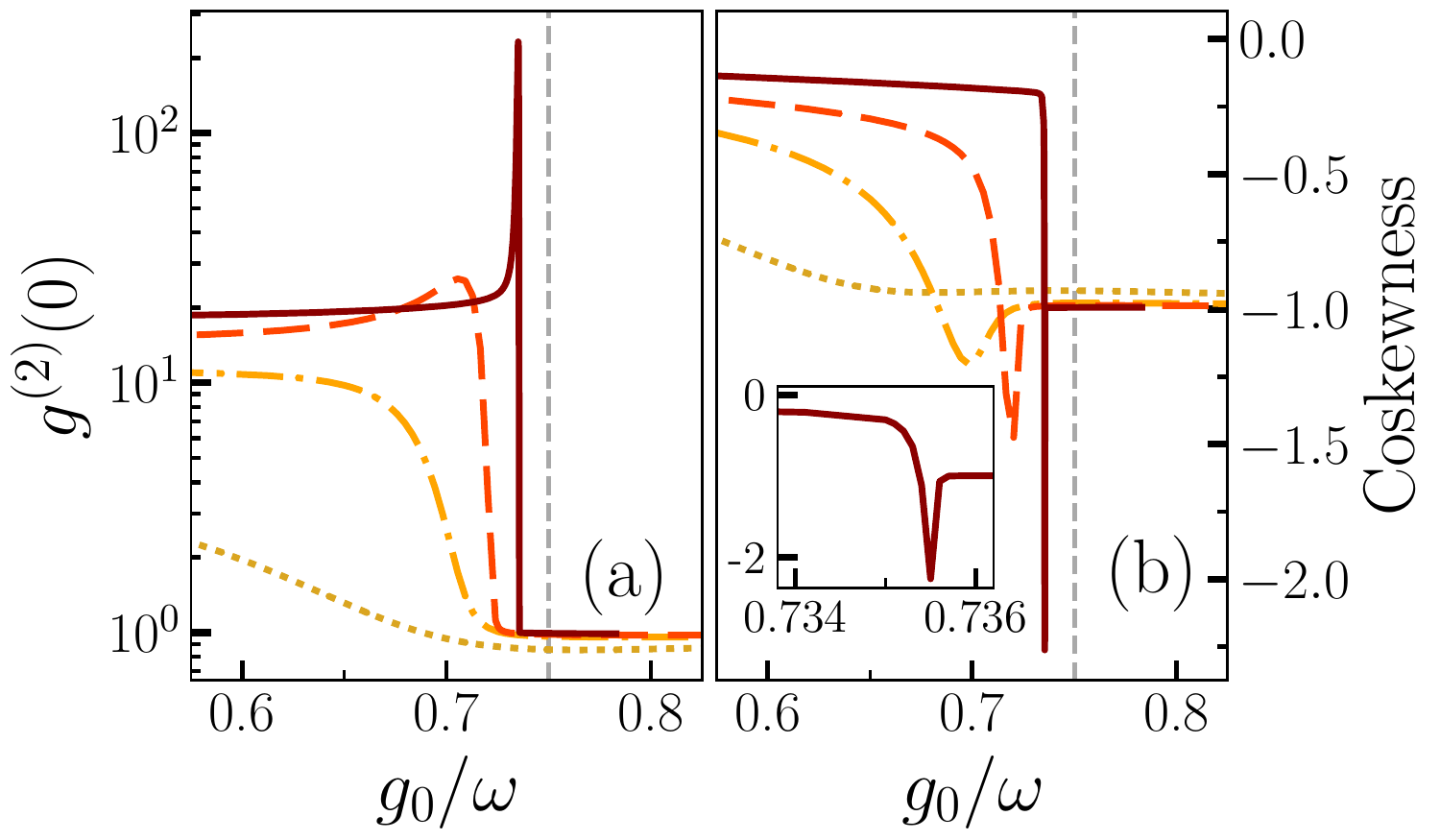}
    \caption{(a) Rescaled equal-time second-order correlation function $g^{(2)}(0)$ and (b) coskweness of the ground state vs the three-body coupling strength $g_0$.
    Inset: zoom of the coskweness for $\eta=10$.
    The first-order transition point predicted by the semiclassical theory is indicated by a vertical dashed line.
    Legend and parameters as in Fig.~\ref{fig:fig_1}.
    }
    \label{fig:fig_2}
\end{figure}

The exact numerical results also reveal a key feature of the transition that is not captured by the semi-classical analysis, namely its non-Gaussian character. By non-Gaussian we mean that the ground state in the vicinity of the transition cannot be described by any superposition of Gaussian states, as predicted by a mean-field approach, even when quantum corrections are included through Bogoliubov theory. As a witness of non-Gaussianity, we consider the coskewness of modes $\ann{a}, \ann{b},\ann{c}$, defined as 
\begin{equation}
\mathcal{C}_{abc} = \frac{\expec{\hat{x}_{a}\hat{x}_{b}\hat{x}_{c}}}{\sqrt{\expec{\hat{x}^2_a}\expec{\hat{x}^2_b}\expec{\hat{x}^2_c}}},
\end{equation}
where by symmetry $\expec{\hat{x}_a} = \expec{\hat{x}_b} = \expec{\hat{x}_c} = 0$. Mean-field theory predicts that in the superradiant phase $\lim_{\eta \to \infty}\mathcal{C}_{abc} = -1$, while $\mathcal{C}_{abc}=0$ in the normal phase \cite{SUPMAT}. Within this approximation, the non-zero value of the coskewness comes from the superposition of coherent states.  When quantum fluctuations are included, this quantity remains bounded and we always find  $\mathcal{C}_{abc} \geq -1$. This stands in sharp contrast with the exact numerical results presented in Fig.~\ref{fig:fig_2} (b), which hint at a divergence of the coskewness at the critical point. To complement the characterization of the ground state we show in Fig.~\ref{fig:fig_2} (a) the single-mode second-order correlation function $g^{(2)}(0) = \frac{\expec{\crea{a}\crea{a}\ann{a}\ann{a}}}{\expec{\crea{a}\ann{a}}^2}$, as a function of the coupling strength. In terms of photon statistics the ground-state undergoes a super-Poissonian-to-Poissonian transition, which does not reveal any non-Gaussian behaviour such as photon antibunching. It is thus only when considering three-mode correlation functions, such as the cowskewness, that the non-Gaussian nature of this superradiant transition becomes apparent.

\paragraph{Implementation with superconducting circuits.---}

Let us now present a scheme to observe this rich phenomenology with current circuit-QED devices~\cite{Gu_review,Blais_review}.  
We generalize the scheme proposed in \cite{FedortchenkoPRA17} and experimentally implemented in \cite{Markovic2018}, which makes use of a spontaneous parametric down conversion (SPDC) to induce an effective USC coupling between two microwave resonators. Here, we consider instead a three-photon SPDC process where a single (pump) photon is down-converted in a photon triplet. This novel quantum process has been recently implemented~\cite{Chang2020} using a superconducting transmission-line resonator grounded through an asymmetric fluxed-pumped SQUID. Non-Gaussian state generation has also been demonstrated~\cite{Agusti2020,Casado2022}. The model of Eq.~\eqref{Eq:Trimer_Hamiltonian} can be implemented using the three-resonator scheme sketched in Fig.~\ref{fig:fig_1}(b). Note that even if similar results might be obtained with a single multimode resonator, a multi-resonator scheme allows for an independent control of the local Kerr nonlinearities. The relevant terms in the SQUID Hamiltonian that generate the required nonlinear processes can be written as~\cite{Chang2020},
\begin{equation}
\label{H_SQ_1}
\hat{H}_{SQ} = \beta_d(t) \sum \chi_k (
\crea{a} + \ann{a} + \crea{b} + \ann{b} + \crea{c} + \ann{c})^k,
\end{equation}
where $\ann{a}$, $\ann{b}$, and $\ann{c}$ are high-quality-factor microwave modes.  
The qubic terms ($k=3$) responsible for the three-photon parametric processes are non-vanishing for an asymmetric SQUID. Most importantly, they can be selectively and simultaneously activated by carefully choosing the drive frequencies~\cite{SUPMAT}. In our scheme, we take the pump to be composed of four harmonic components $\beta_d(t) = \beta_d \sum_i \cos(\omega_i t)$, each of them inducing a third-order parametric interaction. The corresponding Hamiltonian terms can be written as,
\begin{eqnarray}
\label{freq_match}
    \omega_1 &=& \omega_a + \omega_b + \omega_c +\Delta_1
    \longrightarrow \hat{H}_1 = \crea{a}\crea{b}\crea{c} + H.c. \label{freq_match3} \\
    \omega_2 &=& \omega_a + \omega_b - \omega_c +\Delta_2
    \longrightarrow \hat{H}_2 = \crea{a}\crea{b}\ann{c} + H.c.
    \nonumber \\
    \omega_3 &=& \omega_a - \omega_b + \omega_c +\Delta_3
    \longrightarrow \hat{H}_3 = \crea{a}\ann{b}\crea{c} + H.c.
    \nonumber \\
    \omega_4 &=& \omega_a - \omega_b - \omega_c +\Delta_4
    \longrightarrow \hat{H}_4 = \crea{a}\ann{b}\ann{c} + H.c.,
    \nonumber
\end{eqnarray}
where $\omega_a$, $\omega_b$ and $\omega_c$ are the characteristic frequencies of the resonators, while $\Delta_i$ are small detunings. The sum of the four contributions $\hat{H}_i$ reproduces the three-body-interaction term,
\begin{equation}
    \label{H_SQ_2}
    \hat{H}_{SQ} \approx g\sum_i \hat{H}_i =  g \left(\crea{a} + \ann{a} \right)\left(\crea{b}
    + \ann{b} \right) \left(\crea{c} + \ann{c} \right),
\end{equation}
where the pump-induced coupling is given by $g = \beta_d \, \chi_3/2$.
The small detunings $\Delta_i\ll \omega_i$ will establish the frequencies of the bare modes in the effective Hamiltonian, which are resonant for $\Delta_1 = 3 \omega $ and $\Delta_2 = \Delta_3 = \Delta_4 = \omega$. The full target model of Eq.~\eqref{Eq:Trimer_Hamiltonian} is then reproduced in the interaction picture with respect to the Hamiltonian
$H_0 = \left( \omega_a - \omega \right) \crea{a}\ann{a} + \left( \omega_b - \omega \right) \crea{b}\ann{b} + \left( \omega_c - \omega \right) \crea{c}\ann{c}$.
Intrinsic Kerr and cross-Kerr terms are present due to the nonlinearity of the SQUID. However, engineering a local nonlinear element, such as weakly-coupled qubits~\cite{Gu_review,Blais_review}, allows one to make individual Kerr terms dominant and to  tune the size of the non-linearity. In this way, it is possible to explore the finite-frequency scaling with a single sample, although the mode frequencies must be carefully chosen in order to prevent the activation of spurious interaction terms ~\cite{SUPMAT}. Unwanted couplings can become relevant when one of the pump frequencies is  close to two-body resonances $\omega_i\approx\omega_a \pm \omega_b$, or if parametric processes involving non-fundamental modes of the resonators are activated (for instance $\omega_i\approx \omega_a \pm \omega_b  \pm \omega_d$, with  $\omega_d$ the frequency of any higher-order resonator mode).

\begin{figure}
    \centering
    \includegraphics[width=0.49\textwidth]{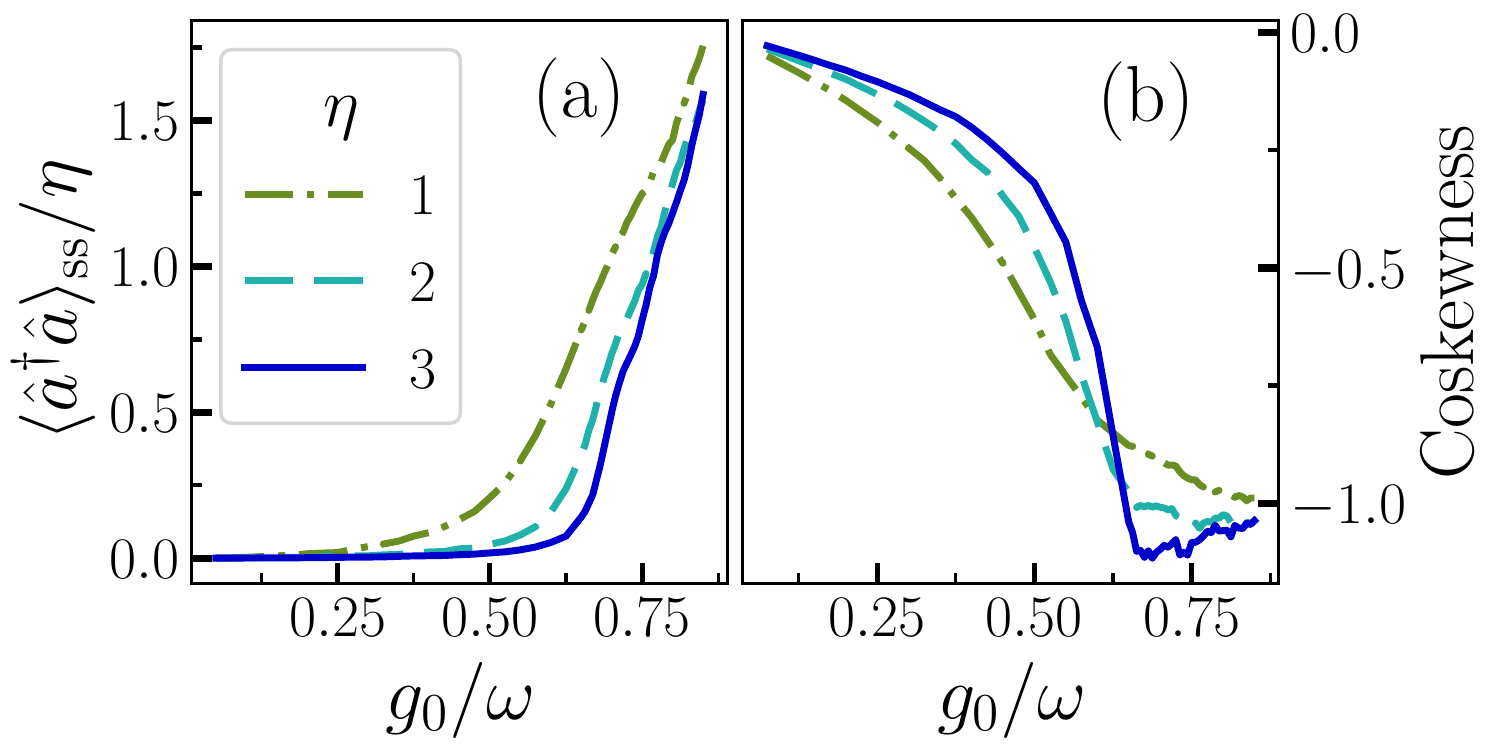}
    \caption{(a) Rescaled photon number  and (b) Coskweness of the steady state vs the three-body coupling strength $g_0$.
    Parameters as in Fig.~\ref{fig:fig_1}, and $\kappa=\omega$. The number of trajectory at each point ensures that observables reached convergence within an error of $5 \%$.
    }
    \label{fig:fig_3}
\end{figure}

\paragraph{Role of dissipation.---}
As any quantum optical setup, the proposed implementation is subject to unavoidable dissipative processes \cite{BreuerBookOpen}. 
In this dissipative context, a phase transition occurs in the steady state, which is reached in the long-time limit under the competition between the unitary dynamics and the loss mechanisms \cite{kessler_generalized_2012,Minganti2018}. We model the driven-dissipative dynamics by the following Lindblad master equation
\begin{equation}\label{Eq:Lindblad1}
\begin{split}
\frac{\de}{\de t}\rhot =  \mathcal L\hat\rho(t)& =-i\left[\hat H,\rhot\right] + \kappa \DD[\hat{a}]\rhot \\ &+ \kappa \DD[\hat{b}]\rhot + \kappa \DD[\hat{c}]\rhot,
\end{split}   
\end{equation}
where $\rhot $ denotes the density matrix of the system. The dissipators describing single-photon losses are defined as $ \DD[\hat{O}]\rhot=\hat{O}\rhot \hat{O}^{\dagger} - (\hat{O}^{\dagger}\hat{O}\rhot+\rhot\hat{O}^{\dagger}\hat{O})/2$
and occur at a rate $\kappa$. Note that in this effective implementation of the USC regime, single-photon losses do not trivially drive the system towards the ground state of the Hamiltonian $\hat{H}$.
Besides unavoidable single-photon losses, the drive can mediate higher-order~\cite{carmichael_statistical2} three-photon dissipative processes. These processes, however, are sub-dominant in the parameter regime we consider and can be safely neglected \cite{Chang2020}. Within the density matrix formalism, the symmetry operators $\hat{S}_i$  are extended to the superoperator level by $\mathcal{S}_i=\hat{S}_i \cdot \hat{S}_i^\dagger$ \cite{BaumgartnerJPA08,Albert2014}. In the present case, the Liouvillian $\LL$ has the same symmetry as $\hat{H}$, namely,
\begin{equation}
[\mathcal{S}_i,\LL] = 0.
\end{equation}
Hence the results of the symmetry analysis extend from the ground state to the steady-state. 

We simulate the dynamics in Eq.~\eqref{Eq:Lindblad1} using a quantum trajectory approach, where the result of the density matrix are obtained by averaging over a large number of realizations of a  stochastic Scr\"odinger equation \cite{SUPMAT}.
The steady state values are then obtained by evolving the results for sufficiently long times, and the convergence of the observables is assured within $5\%$. Three factors make the simulation demanding for large $\eta$ values: (i) Growing photon numbers require a large cutoff. (ii) The non-Gaussian nature of the state implies long tails in the photon number distribution. (iii) The state is extremely entropic at the phase transition, requiring a large trajectory sample to obtain non-noisy data.
In Fig.~\ref{fig:fig_3}(a), we show the average photon number as a function of the coupling $g$.
Similarly to the nondissipative case, we observe a sharper change in the photon number as the parameter $\eta$ increases. Notice also that the transition point is slightly shifted with respect to the nondissipative case. Finally, in Fig.~\ref{fig:fig_3}(b) we show that also in the dissipative case the coskewness drops below $-1$, signalling the non-Gaussian nature of the transition. 
We thus confirm that the proposed implementation allows witnessing the unconventional properties of superradiant phase transitions induced by ultrastrong three-body coupling.

\paragraph{Conclusions.---}
We have demonstrated that a trimer of nonlinear oscillators coupled via a 3-body terms exhibits a superradiant phase transition in the USC regime. In contrast with the usual two-body dipolar coupling, the 3-body Hamiltonian leads to large non-Gaussian quantum fluctuations, as witnessed by a diverging coskewness in the vicinity of the transition. Exact numerical simulations show that these features, which are captured neither by semiclassical analysis nor by Bogoliubov approach, are robust to dissipation and could be observed with the proposed circuit-QED scheme. Our results demonstrate the theoretical and experimental relevance of three-body couplings in open quantum-optical systems. This class of models can lead to the observation of a novel quantum phenomenology related to the ultrastrong coupling regime~\cite{Forn-Diaz2019,Frisk-Kockum2019,LeBoite2020}, and to applications in quantum simulations and quantum-information processing~\cite{Lieu2020}.

\begin{acknowledgments}
A.L.B. and S.F. acknowledge support from CNRS via the International Emerging Action project DDMolPol (203844). L. G. was supported by the Austrian Academy of Sciences (\"OAW) and by the Austrian Science Fund (FWF) through Grant No. M3214 (ASYMM-LM).
\end{acknowledgments}

\bibliographystyle{apsrev4-1}
\bibliography{References.bib}

\clearpage
\widetext
\begin{center}
\textbf{\huge Supplemental Material}
\end{center}

\setcounter{equation}{0}
\setcounter{figure}{0}
\setcounter{table}{0}
\setcounter{page}{1}
\setcounter{lemma}{0}

\makeatletter

\renewcommand{\theequation}{S\arabic{equation}}
\renewcommand{\thelemma}{S\arabic{lemma}}
\renewcommand{\thefigure}{S\arabic{figure}}
\renewcommand{\bibnumfmt}[1]{[S#1]}
\renewcommand{\citenumfont}[1]{S#1}
\renewcommand{\thepage}{S\arabic{page}}

\section{Effective implementation}

Let us now provide more details on the proposed effective implementation that can reproduce the desired model. In particular, we will discuss the requirements that the driving fields and the resonator mode structure must fulfill in order to avoid the activation of unwanted coupling terms. The full quantum model for the three-resonator setup sketched in Fig.~\ref{fig:fig_1}(b) can be written as,
\begin{equation}
\label{Eq:Full}
    \hat{H} = \sum_n \left\{ \omega_a^{(n)} \crea{a_n}\ann{a_n} + \omega_b^{(n)} \crea{b_n}\ann{b_n} + \omega_c^{(n)} \crea{c_n}\ann{c_n} \right\} + \hat{H}_{kerr} + \hat{H}_{SQ},
\end{equation}
where,
\begin{equation}
\hat{H}_{kerr}= \sum_n U^{(n)} \left(\hat{a}_n^{\dag 2}\ann{a_n}^2 + 
\hat{b}_n^{\dag 2}\ann{b_n}^2 + \hat{c}_n^{\dag 2}\ann{c_n}^2\right)
\end{equation}
We defined the modes $\ann{a_n}$ of the resonator $a$, with resonant frequency $\omega_a^{(n)}$ and Kerr strength $U^{(n)}$, and similarly for the resonators $b$ and $c$. Notice that $n=0$ corresponds to the fundamental mode considered in the main text $\ann{a_0}\equiv\ann{a}$, where we dropped the suffix for the sake of simplicity. The full SQUID Hamiltonian then reads
\begin{equation}
\label{sup_H_SQ_full}
\hat{H}_{SQ} = \beta_d \left(\sum_i \cos(\omega_i t) \right)\sum_k \chi_k \left[
\sum_n \left( \crea{a_n} + \ann{a_n} + \crea{b_n} + \ann{b_n} + \crea{c_n} + \ann{c_n} \right)
\right]^k.
\end{equation}
In the interaction picture defined by the harmonic part of the free Hamiltonian, the annihilation/creation operators rotate each at the corresponding resonant frequency $\ann{a_n}(t) = \ann{a_n}  e^{i\omega_a^{(n) t}}$, and fast oscillating terms can be neglected by rotating-wave approximation given that $\beta_d \chi_k$ is much smaller than all mode frequencies. As a first condition, we take the three resonators to have different fundamental frequencies $\omega_a^{(0)} \neq \omega_b^{(0)} \neq \omega_c^{(0)}$, in such a way that intrinsic coupling terms are off-resonant and so photon transfer is possible only when mediated by the drivings. 
We have then to select the frequencies $\omega_i$ of the harmonic components of the driving field, in such a way that the desired processes are resonant and that no spurious terms are activated. In order to reproduce the three-photon coupling we need four components,
\begin{eqnarray}
\label{sup_freq_match}
    \omega_1 &=& \omega_a^{(0)} + \omega_b^{(0)} + \omega_c^{(0)} +\Delta_1
    \quad \longrightarrow \quad \hat{H}_1 = 
    \frac{\beta_d \chi_3}{2} \left(\crea{a_0}\crea{b_0}\crea{c_0} + H.c. \right) \label{freq_match3} \\
    \omega_2 &=& \omega_a^{(0)} + \omega_b^{(0)} - \omega_c^{(0)} +\Delta_2
    \quad \longrightarrow \quad \hat{H}_2 = 
    \frac{\beta_d \chi_3}{2} \left( \crea{a_0}\crea{b_0}\ann{c_0} + H.c.\right)
    \nonumber \\
    \omega_3 &=& \omega_a^{(0)} - \omega_b^{(0)} + \omega_c^{(0)} +\Delta_3
    \quad \longrightarrow \quad \hat{H}_3 = 
    \frac{\beta_d \chi_3}{2} \left( \crea{a_0}\ann{b_0}\crea{c_0} + H.c.\right)
    \nonumber \\
    \omega_4 &=& \omega_a^{(0)} - \omega_b^{(0)} - \omega_c^{(0)} +\Delta_4
    \quad \longrightarrow \quad \hat{H}_4 = 
    \frac{\beta_d \chi_3}{2} \left( \ann{a_0}\crea{b_0}\crea{c_0} + H.c.\right).
    \nonumber
\end{eqnarray}
 As explained in the main text, the small detunings $\Delta_i\ll \omega_i$ will establish the frequency $\omega$ of the bare modes in the effective Hamiltonian  $\Delta_1 = 3 \omega $, and $\Delta_2 = \Delta_3 = \Delta_4 = \omega$. 
 Furthermore, the driving frequencies must be significantly detuned from all resonances that induce energy transfer or parametric couplings between  fundamental and higher modes. In particular, two kind of unwanted terms might be activated: Two photon processes such as for example $\ann{a_0}\crea{b_1}$, non-negligible for some $\omega_i \sim \omega_b^{(1)} - \omega_a^{(0)}$, and three-photon processes such as $\crea{a_0}\crea{c_1}\crea{c_1}$, which can be activated when $\omega_i \sim 2\omega_c^{(1)} + \omega_a^{(0)}$. In order for these processes to be negligible, all driving frequencies $\omega_i$ must be detuned by an amount $\delta$ which is large with respect to the effective coupling strength. 

For a given choice of mode frequencies, we can numerically check the detuning with respect to all undesired resonances. Let us provide an example of a suitable set of parameters which is in line with the experimental implementation presented in \cite{Chang2020}: $(\omega_a^{(0)},\omega_b^{(0)},\omega_c^{(0)}) = 2\pi(7.6,6.2,4.2)$GHz, and $(\omega_a^{(1)},\omega_b^{(1)},\omega_c^{(1)}) = 2\pi(11.4,  9.3,  6.3)$GHz. For this set of frequencies, and chosing the drive tones as in Eq.~\eqref{sup_freq_match}, we find that the closest unwanted resonance is given by 
$\delta = \omega_1 - \omega_a^{(0)} - \omega_c^{(0)} - \omega_c^{(1)} = 2\pi\ 0.1$GHz, and other three-photon processes with the same detuning $\delta$. We have to compare this detuning with the strength $g = \frac{\beta_d\chi_3}{2}$of the process which, in order to observe the transition described in the main text, must reach the value such that 
$ \lambda = \sqrt{{2}/ ({\omega U})}g = {3}/{(2\sqrt{2})}$, and so $g ={3}\sqrt{\omega U}/{4}$. If we fix now the frequency of the effective model $\omega =2\pi\ 10$MHz and $U = \omega/10$ (corresponding to $\eta = 10$), we have that $g\sim 2\pi\ 2.4 $. Accordingly, the ratio $g/\delta\sim 2.4\time 10^{-2}$ is much smaller than 1, and these processes are safely negligible with standard physical parameters~\cite{Markovic2018,Chang2020}. 


\section{Semi-classical treatment of the trimer Hamiltonian}

\subsection{Mean-field approximation}
We start from the trimer Hamiltonian:
\begin{align}
\label{TrimHam}
\Hop=\omega (\adag\aop + \bdag\bop + \cdag\cop) + \frac{g_0}{\sqrt{\eta}} (\adag+\aop)(\bdag+\bop)(\cdag+\cop) + \frac{U_0}{\eta} (\aop^{\dag 2}\aop^2 + \bop^{\dag 2}\bop^2 + \cop^{2\dag}\cop^2)
\end{align}
We will first study this Hamiltonian with a classical approximation. We assume that each field can be put in a coherent state: $\aop\rightarrow\alpha$, $\bop\rightarrow\beta$, $\cop\rightarrow\gamma$; the consistency of this hypothesis will be studied at the end of this section. The Hamiltonian then reduces to a classical potential:

\begin{equation}
 H(\alpha,\beta,\gamma)=\omega(\lvert\alpha\rvert^2+\lvert\beta\rvert^2+\lvert\gamma\rvert^2)+\frac{g_0}{\sqrt{\eta}}(\alpha+\alpha^*)(\beta+\beta^*)(\gamma+\gamma^*)+\frac{U_0}{\eta}(\lvert\alpha\rvert^4+\lvert\beta\rvert^4+\lvert\gamma\rvert^4)
\end{equation}

The goal then is to find the minima and maxima of this function, which satisfy $\partial H/\partial\alpha=\partial H/\partial\beta=\partial H/\partial\gamma=0$. We find nine solutions: $\alpha=\beta=\gamma=0$, $\alpha=\beta=-\gamma=X_\pm$, $\alpha=-\beta=\gamma=X_\pm$, $-\alpha=\beta=\gamma=X_\pm$, and $\alpha=\beta=\gamma=-X_\pm$, with 
\begin{equation}
X_\pm=\sqrt{\frac{\omega \eta}{2 U_0}}(\lambda\pm\sqrt{\lambda^2-1})=\sqrt{\frac{\omega \eta}{2 U_0}}f_\pm(\lambda)
\end{equation}
where we have defined:
\begin{equation}
    \lambda=\sqrt{\frac{2}{\omega U_0}}g_0=\sqrt{\frac{2}{\omega U}}g.
\end{equation}
Note that these solutions exist only for $\lambda>1$.  
The eight solutions can be grouped in two sets of four degenerate solutions, with a (semiclassical) energy:

\begin{equation}
    E_\pm(\lambda)=\frac{3\omega^2\eta}{4 U_0}\left(2f_\pm(\lambda)^2-\frac{8}{3}\lambda f_\pm(\lambda)^3+f_\pm(\lambda)^4\right)
\end{equation}

Importantly, we have $E_-\geq0$ and $E_-\geq E_+$. To summarize, for $\lambda<1$, we have a semiclassical potential with a single minimum which corresponds to the vacuum. For $\lambda>1$, the potential has five asymetric minima and four maxima. The five minima corresponds to the vacuum and the four solutions $X_+$, and have energy $E=0$ and $E=E_+$ respectively. The maxima corresponds to the four solutions $X_-$, with an energy $E=E_-$. With this analysis, we can already predict that the solutions $X_+$ will always be unstable. 

To better visualize the structure of the potential, let us see what happens if we move along the line $\alpha=\beta=\gamma$. Starting from $\alpha=0$, the potential increases until we reach $\alpha=X_-$; it then decreases until $\alpha=X_+$, then goes up again. We obtain the same picture if we follow the line $\alpha=\beta=-\gamma$ and so on. At first, for $\lambda$ above $1$, the four superradiant states can then exist as metastable states, living in wells of typical width $X_+-X_-$, and with a potential barrier of height $\lvert E_+-E_-\rvert$. As $\lambda$ increases, the depth of the four wells increases. It is straightforward to show that, at $\lambda=\frac{3}{2\sqrt{2}}$, $E_+$ becomes smaller than the vacuum energy $0$. At this point, the four superradiant solutions become the new (degenerate) ground state, and a first-order phase transition takes place. In the following, to lighten up notations, we will rewrite the position of the local minima as $\bar{X}=-X_-$; this is the notation we have used in the main text.\\

\subsection{Quantum fluctuations}
 The treatment above predicts that superradiant states become energetically favored above a certain threshold for the coupling. However, quantum fluctuations can also induce tunnelling between different states and change their stability. We will now estimate the importance of these fluctuations by a quadratic expansion around the semi-classical solution.\\
 
 \paragraph{Quadratic expansion and Bogoliubov transformation.---}
 Let us consider the phase corresponding to $\alpha=\beta=\gamma=\bar{X}$ (the other three phases will have the same stability, by symmetry). We can then decompose the bosonic field into its mean-field value, plus quantum fluctuations:
 $$\aop\rightarrow\bar{X}+\hat{\mu}, \hspace{5 pt}\bop\rightarrow\bar{X}+\hat{\nu},\hspace{5pt}\cop\rightarrow\bar{X}+\hat{\zeta}$$ 
 We can then develop the Hamiltonian in various orders of perturbation. The zero-order term is a constant, and corresponds to the semiclassical energy $E_+$. The first-order term is zero, since we are developing around an energy extremum. The second-order term is: 
\begin{equation}
 \Hop^{(2)}=\frac{\omega}{4}\left(1+f_+(\lambda)^2 \right) \left( \pop_\mu^2+\pop_\nu^2+\pop_\zeta^2 \right)+\frac{\omega}{4} \left(1+3f_+(\lambda)^2 \right) \left(\xop_\mu^2+\xop_\nu^2+\xop_\zeta^2 \right)-\omega\lambda f_+(\lambda)\left[ \xop_\mu\xop_\nu+\xop_\nu\xop_\zeta+\xop_\zeta\xop_\mu \right]
 \label{H2_first}
\end{equation}
where we have defined the quadratures $\xop_\mu=\hat{\mu}+\hat{\mu}^\dag$, and so on. This Hamiltonian can be diagonalized with a Bogoliubov transformation. We define new fields $\uop$, $\yop$ and $\zop$ as:

\begin{equation}
    \begin{bmatrix}    \xop_\mu \\ \xop_\nu\\ \xop_\zeta    \end{bmatrix} =\frac{1}{\sqrt{3}} \left(\frac{1+f_+^2}{1+3f_+^2}\right)^{1/4} \begin{bmatrix}
    1&1&1\\ -\frac{1+\sqrt{3}}{2} &1&\frac{\sqrt{3}-1}{2}\\\frac{\sqrt{3}-1}{2}&1&  -\frac{1+\sqrt{3}}{2}
    \end{bmatrix} \begin{bmatrix}    \uop \\ \yop\\ \zop    \end{bmatrix}
    \label{Bogo_suprad}
\end{equation}

The conjugate variables $\pop_u$, $\pop_y$, $\pop_z$ are instead defined by:
\begin{equation}
    \begin{bmatrix}    \pop_\mu \\ \pop_\nu\\ \pop_\zeta    \end{bmatrix} =\frac{1}{\sqrt{3}} \left(\frac{1+3f_+^2}{1+f_+^2}\right)^{1/4} \begin{bmatrix}
    1&1&1\\ -\frac{1+\sqrt{3}}{2} &1&\frac{\sqrt{3}-1}{2}\\\frac{\sqrt{3}-1}{2}&1&  -\frac{1+\sqrt{3}}{2}
    \end{bmatrix} \begin{bmatrix}    \pop_u \\ \pop_y\\ \pop_z    \end{bmatrix}
\end{equation}

These operators define \textit{collective} excitations of the three resonators. In terms of these new degrees of freedom, the Hamiltonian above reads:

\begin{equation}
    \Hop^{(2)}=\frac{\tilde{\omega}}{4}[\pop_u^2+\pop_y^2+\pop_z^2]+\frac{\tilde{\omega}}{4}\left[\uop^2\left(1+\frac{\tilde{\lambda}}{2}\right)+\zop^2\left(1+\frac{\tilde{\lambda}}{2}\right)+\yop^2(1-\tilde{\lambda})\right]
    \label{H2_final}
\end{equation}
Where the renormalized coupling $\tilde{\lambda}$ and frequency $\tilde{\omega}$ are defined as:
\begin{align}
(1-\tilde{\lambda})&=\sqrt{\lambda^2-1}\left[\frac{\lambda+\sqrt{\lambda^2-1}}{3\lambda\sqrt{\lambda^2-1}+3\lambda^2-1}\right]\\
\tilde{\omega}&=\omega\sqrt{(1+f_+^2)(1+3f_+^2)}
\end{align}
One can readily check that, for $\lambda>1$, all collective modes are stable and have bounded fluctuations. At $\lambda=\tilde{\lambda}=1$,  however, the $y$ polariton becomes unstable. This is compatible with the mean-field treatment, which predicted that superradiant states exist only for $\lambda>1$ in the first place.

\paragraph{Analysis of fluctuations.---}
Let us study in details the quadrature fluctuations predicted by the previous analysis.
In the normal phase, we have the usual vacuum fluctuations $\langle\xop_i^2\rangle=\langle\pop_i^2\rangle=1$, and $\langle\xop_i\xop_j\rangle=0$, for $i={\mu,\nu,\zeta}$. 
In the superradiant phase, combining Eqs.~\eqref{Bogo_suprad} and \eqref{H2_final}, we find the following values:  $\langle\pop_y^2\rangle=\sqrt{1-\tilde{\lambda}}$, $\langle\uop\yop\rangle=\langle\zop\yop\rangle=\langle\uop\zop\rangle=\langle\pop_u\pop_y\rangle=\langle\pop_z\pop_y\rangle=\langle\pop_u\pop_z\rangle=0$. With this, we find the following expressions for the fluctuations of the $u$, $y$, $z$ quadratures:

\begin{align}
\label{fluctu_suprad}
    \langle\uop^2\rangle=\langle\zop^2\rangle&=\frac{1}{\sqrt{1+\tilde{\lambda}/2}}\\
    \langle\yop^2\rangle&=\frac{1}{\sqrt{1-\tilde{\lambda}}}\\
    \langle\uop\yop\rangle&=\langle\zop\yop\rangle=\langle\uop\zop\rangle=0\\
    \langle\xop_\mu^2\rangle&=\langle\xop_\nu^2\rangle=\langle\xop_\zeta^2\rangle=\frac{1}{3}\left(\frac{1+f_+^2}{1+3f_+^2}\right)^{1/2}\Big(2\langle\zop^2\rangle+\langle\yop^2\rangle\Big)\label{onefluctu_sup}\\ 
    \langle\xop_\mu\xop_\nu\rangle&=\langle\xop_\nu\xop_\zeta\rangle=\langle\xop_\zeta\xop_\mu\rangle=\frac{1}{3}\left(\frac{1+f_+^2}{1+3f_+^2}\right)^{1/2}\Big(\langle\yop^2\rangle-\langle\zop^2\rangle\Big)\\
    \langle\xop_\mu\xop_\nu\xop_\zeta\rangle&=0
    \label{trifluctu_sup}
\end{align}
and for the $p$ quadratures:
\begin{align}
    \langle\pop_u^2\rangle&=\langle\pop_z^2\rangle=\sqrt{1+\tilde{\lambda}/2}\\
    \langle\pop_y^2\rangle&=\sqrt{1-\tilde{\lambda}}\\ 
    \langle\pop_z\pop_y\rangle&=\langle\pop_u\pop_z\rangle=\langle\pop_u\pop_y\rangle=0\\
    \langle\pop_\mu^2\rangle&=\langle\pop_\nu^2\rangle=\langle\pop_\zeta^2\rangle=\frac{1}{3}\left(\frac{1+3f_+^2}{1+f_+^2}\right)^{1/2}\Big(2\langle\pop_z^2\rangle+\langle\pop_y^2\rangle\Big)\\
    \langle\pop_\mu\pop_\nu\rangle&=\langle\pop_\nu\pop_\zeta\rangle=\langle\pop_\zeta\pop_\mu\rangle=\frac{1}{3}\left(\frac{1+3f_+^2}{1+f_+^2}\right)^{1/2}\Big(\langle\pop_y^2\rangle-\langle\pop_z^2\rangle\Big)\\
    \langle\pop_\mu\pop_\nu\pop_\zeta\rangle&=0
\end{align}

The fluctuations, which diverge near $\tilde{\lambda}=1$ (which means $\lambda=1$), are reminiscent to what we obtain in models such as the Dicke model \cite{Kirton2019} or the Bose-Hubbard dimer \cite{Felicetti2020}. There is, however, a key difference: in Dicke-like models, the appearence of the superradiant phase (which is associated with diverging fluctuations) coincide with the phase transition. Here, at $\lambda=1$, the superradiant phases are still high-energy phases, and the ground state is still centered around $\alpha=\beta=\gamma=0$. The phase transition occurs only for $\lambda={3}/{(2\sqrt{2})}$; at this point, the semi-classical treatment we have presented predicts that the superradiant phases are already stabilized. Indeed, according to the formulas above, both the on-site fluctuations $\langle \xop_i^2 \rangle$ and the cross-sites correlations  $\langle\xop_i\xop_j\rangle$ are finite for $\lambda={3}/{(2\sqrt{2})}$ .\\

All of these expressions have been obtained by a development around the solution $\alpha=\gamma=\beta=\bar{X}$. If we develop the solution around, say, $\alpha=\beta=-\gamma=-\bar{X}$, then we will obtain the Hamilonian \eqref{H2_first}, up to a transformation $\aop\rightarrow-\aop$ and $\bop\rightarrow-\bop$. The on-site fluctuations $\langle\xop_i^2\rangle$ will remain the same, but the cross-site correlations $\langle\xop_i\xop_j\rangle$ will change. We will observe noise reduction for the following quadratures: $\xop_\mu-\xop_\nu$, $\xop_\nu+\xop_\zeta$, $\xop_\mu+\xop_\zeta$. Similarly, for the other two solutions, we will find the same squeezing amount, but different squeezing directions. In the following, we will refer to the four superradiant states as $\ket{G_{+++}}$, $\ket{G_{+--}}$, $\ket{G_{-+-}}$, and $\ket{G_{--+}}$.\\

\subsection{Stability of the superradiant state}
The analysis above allows us to make a further comment on the stability of the superradiant states. Just above $\lambda=1$, superradiant state can in principle exist, but quantum fluctuations can still destabilize the phase.
Near the point $\lambda=1$, we have $\langle\yop^2\rangle\sim(\tilde{\lambda}-1)^{-1/2}\sim(\lambda-1)^{-1/4}$, while $\langle\uop^2\rangle$ and $\langle\zop^2\rangle$ remain bounded and of order $1$, because the average values $\langle\uop\rangle$, $\langle\yop\rangle$ and $\langle\zop\rangle$ are zero everywhere. The quantum state will therefore be dominated by the fluctuations of the $\yop$ polariton, and we have $\langle\xop_\mu^2\rangle\sim\langle\xop_\nu^2\rangle\sim\langle\xop_\zeta^2\rangle\sim\langle\yop^2\rangle$. We can now compare these fluctuations to the width of the well:
\begin{equation}
   \frac{\langle\xop_\mu^2\rangle}{(X_+-X_-)^2}\sim \frac{\langle\xop_\nu^2\rangle}{(X_+-X_-)^2}\sim \frac{\langle\xop_\zeta^2\rangle}{(X_+-X_-)^2}\sim\frac{U_0}{\omega\eta(\lambda-1)^{5/4}}
\end{equation}
For $\lambda\lessapprox 1 +  l= 1+\left(\frac{U_0}{\eta\omega}\right)^{4/5}$, where $l = \left(\frac{U_0}{\eta\omega}\right)^{4/5}$, the field fluctuations are of the same order of magnitude that the width of the potential well. Therefore, the quantum fluctuations can kick the system out of the local minima. Only for $\lambda\gtrapprox 1+\left(\frac{U_0}{\eta\omega}\right)^{4/5}$ become the superradiant states truly well-defined and (meta)stable. Alternatively, we may also compare the excitation energy with the potential barrier. The excitation energy will scale like $\omega(1-\lambda^{1/4})$, while the potential barrier gives $E_+-E_-\sim \frac{\omega^2\eta}{U_0}(1-\lambda)^{3/2}$. Again, we find that the excitation energy becomes smaller than the barrier for $\lambda\gtrapprox 1+\left(\frac{U_0}{\eta\omega}\right)^{4/5}$; meaning that only above this point, we can supress tunelling to the vacuum state, and stabilize the superradiant state.\\

As of the normal phase, it will remain stable for most values of $g$. However, for very large values of $\lambda$, we have $E_-\rightarrow 0$; as a consequence, the potential barrier isolating the vacuum from the superradiant phase vanishes, and quantum fluctuations drive the system out of the vacuum. We expect this will occur when the excitation energy in the normal phase becomes comparable with the barrier, \textit{i.e.}, for $\omega\sim E_-$, which gives $\lambda\sim \sqrt{\omega \eta/U_0}$, or equivalently $g_0\sim \omega\sqrt{\eta}$. A similar order of magnitude can be obtained with the following reasoning: if we keep only the quadratic term in the Hamiltonian \eqref{TrimHam}, we predict fluctuations $\langle \xop_a^2 \rangle=O(1)$. Hence, the quadratic potential $\omega (\adag\aop+..)$ will be of order $\omega$, and the trimer interaction term will be of order $\frac{g_0}{\sqrt{\eta}}\xop^3\sim\frac{g_0}{\sqrt{\eta}}$.  
Hence, for $g_0\ll \sqrt{\eta}$, the interaction term will be negligible; when $g_0\sim\sqrt{\eta}$, the interaction becomes comparable to the quadratic potential, and can destabilize the normal phase. This means that this mean-field treatment predicts that the normal phase becomes unstable well after the ground state becomes superradiant.\\

To summarize everything: for $\lambda<1$, the potential has a single minimum which corresponds to the vacuum. For $1<\lambda\lessapprox 1+(U_0/\eta\omega)^{4/5}$, four degenerate minima appear; however, the local fluctuations are still strong enough to drive the system out of these minima. For $1+(U_0/\eta\omega)^{4/5}\lessapprox\lambda<\frac{3}{2\sqrt{2}}$, the superradiant state becomes metastable. However, its energy is still larger than the vacuum state energy. For $\lambda=\frac{3}{2\sqrt{2}}$, the superradiant and vacuum states become degenerate, and a phase transition takes place. Note that at this point, there is a large potential barrier between the two states, and tunneling between the vacuum and superradiant states is still supressed: the transition is first-order. For $\lambda>\frac{3}{2\sqrt{2}}$, the ground state is now superradiant, however the vacuum remains a minimum of potential for all values of $\lambda$. For most values of $\lambda$, this minimum of potential remains deep enough to confine the field: the vacuum is still metastable. Only for $\lambda\sim \sqrt{\omega \eta/U_0}$, the fluctuations induced by the interaction become strong enough to destabilize the vacuum. This semi-classical analysis seems to capture correctly the location of the critical point, as well as the mean number of excitations in the superradiant phase (see Fig.\ref{fig:fig_1} in the main text). However, as we will shortly show, it fails to capture the divergence of the coskewness at the critical point.\\


\subsection{Coskewness}

\paragraph{Normal phase} We will now study the third-order moments, starting with the normal phase. As long as we remain in the vicinity of the vacuum state, the three-body coupling will only act as a perturbation. Using standard perturbation theory at first order, we find that the ground state will be given by $\ket{000}-\frac{g_0}{3\omega\sqrt{\eta}}\ket{111}$. We can immediately infer the skewness of each quadrature fluctuation:

\begin{align}
    \langle\xop_a^3\rangle=0\\
    \langle\xop_a^2\xop_b\rangle=0\\
     \langle(\xop_a\xop_b\xop_c)\rangle&=\frac{-g_0}{3\omega\sqrt{\eta}}\\
\end{align}
    
And the same expressions are obtained by permuting $a$, $b$ and $c$. Hence, only the three-mode phase space distribution will be skewed. This gives:

\begin{align}
    \mathcal{C}_{abc} = \frac{\expec{\hat{x}_{a}\hat{x}_{b}\hat{x}_{c}}}{\sqrt{\expec{\hat{x}^2_a}\expec{\hat{x}^2_b}\expec{\hat{x}^2_c}}}\propto-\frac{g_0}{\omega\sqrt{\eta}}
\end{align}

Hence, the coskewness in the normal phase is negative, and tends to zero when $\eta$ tends to infinity. This is indeed what we observe on Fig.\ref{fig:fig_2} of the main text. Additionally, we can also derive:

\begin{equation}
    \left\langle\left(\frac{\xop_a+\xop_b+\xop_c}{\sqrt{3}}\right)^3\right\rangle=\left\langle\left(\frac{\xop_a-\xop_b-\xop_c}{\sqrt{3}}\right)^3\right\rangle=\left\langle\left(\frac{-\xop_a-\xop_b+\xop_c}{\sqrt{3}}\right)^3\right\rangle=\left\langle\left(\frac{-\xop_a+\xop_b-\xop_c}{\sqrt{3}}\right)^3\right\rangle=\frac{-g_0}{3\omega\sqrt{\eta}}
\end{equation}

This means that the distribution of the four quadratures above is biaised towards negative values. The directions towards which the distribution is biaised corresponds precisely to the four possible directions of displacement in the superradiant phase. This is very similar to Dicke-like transitions, in which the distribution in phase space prior to the transition is distorted along the axis of displacement in the superradiant phase. Here however the transition is first-order, which means that we expect an abrupt transition from a skewed distribution centered around the vacuum to a four-fold displaced distribution.\\

\paragraph{Superradiant phase}
Let us now look at the coskewness in the superradiant phase. 
In the thermodynamic limit, semiclassical and Gaussian theories leads to a four-fold degenerate groundstate composed of any superposition of four displaced squeezed states. However, numerical simulations show that for physical (finite) values of $\eta$ the gap decreases in the proximity of the expected critical point but is never exactly vanishing. The ground state is then well approximated by the superposition of the four superradiant states which is in the same parity subspace of the groundstate in the normal phase (the vacuum). We show in the following that this leads to accurate predictions for the value of the coskewness deep in the superradiant phase but, as expected, it cannot explain the divergence in proximity of the critical point.
If we neglect quantum fluctuations and consider coherent states, the ground state would then be given by,
\begin{align}
\ket{G} = \frac{1}{2}\left( \ket{\bar{X},\bar{X},\bar{X}} +\ket{-\bar{X},-\bar{X},\bar{X}} + \ket{-\bar{X},\bar{X},-\bar{X}} + \ket{\bar{X},-\bar{X},-\bar{X}}\right) .
\end{align}

The three-body correlations are given by $$\bra{G} \xop_a \xop_b \xop_c \ket{G} =  8 \bar{X}^3.$$
The results is given by the sum of four equivalent contributions of the diagonal terms. The non-diagonal terms of the form $\bra{\bar{X},\bar{X},\bar{X}} \xop_a \xop_b \xop_c \ket{-\bar{X},-\bar{X},\bar{X}}$ all cancel (indeed, $\bra{\bar{X},\bar{X},\bar{X}} \xop_a \xop_b \xop_c \ket{-\bar{X},-\bar{X},\bar{X}}=\bra{\bar{X}}\xop_a\ket{-\bar{X}}\bra{\bar{X}}\xop_b\ket{-\bar{X}}\bra{\bar{X}}\xop_c\ket{\bar{X}}$, and $\bra{\bar{X}}\xop_a\ket{-\bar{X}}=\bra{\bar{X}}\aop\ket{-\bar{X}}+\bra{\bar{X}}\adag\ket{-\bar{X}}=0$).\\

The square of each quadrature $\xop$ will give,
\begin{align}
\nonumber
\bra{G} \xop_a^2 \ket{G}&= \frac{1}{4}\left[ 2(\bra{\bar{X}}\xop_a^2\ket{\bar{X}}+\bra{-\bar{X}}\xop_a^2\ket{-\bar{X}})(1+\bracket{\bar{X}}{-\bar{X}}^2) +8\bra{\bar{X}}\xop_a^2\ket{-\bar{X}}\bracket{\bar{X}}{-\bar{X}}\right]\\
&= \frac{1}{4}\left[ 4(4 \bar{X}^2 + 1) + 4(4\bar{X}^2+3) \bracket{\bar{X}}{-\bar{X}}^2 \right],
\end{align}
where the first and second terms are given by the sum of all diagonal and off-diagonal contributions, respectively. Finally, we find that all the linear terms  $\bra{G} \xop_i \ket{G}$, cancel. This finally yields the expression of the coskewness:
\begin{equation}\label{Eq:coskewCoherent}
\mathcal{C}_{abc} = \frac{ 8 \bar{X}^3}{ \left(4\bar{X}^2 + 1 +(4\bar{X}^2+3)\ e^{-4 \bar{X}^2} \right)^{3/2} }.
\end{equation}
 Deep into the symmetry-broken phase, we recover $\lim_{\eta\to\infty} \mathcal{C}_{abc} = -1$ as found numerically. However, this treatment also predicts that we always have $\mathcal{C}_{abc}>-1$. Therefore, even by taking into account the superposition of four coherent states, the semi-classical analysis cannot account for the divergence of the coskewness at the critical point. 
 
 So far we have focused on the semiclassical theory. Let us now show that including Gaussian fluctuations cannot affect significantly the value of $\mathcal{C}_{abc}$. For instance, let us consider  the term $\bra{G_{+++}}\xop_a\xop_b\xop_c\ket{G_{+++}}$ which appears in the expression of $\mathcal{C}_{abc}$. We can decompose it as:
 $$\bra{G_{+++}}\xop_a\xop_b\xop_c\ket{G_{+++}}=8\bar{X}^3+12\bar{X}^2\langle\xop_\mu\rangle+6\hat{X}\langle\xop_\mu\xop_\nu\rangle+\langle\xop_\mu\xop_\nu\xop_\zeta\rangle=8\bar{X}^3+2\bar{X}\left(\frac{1+f_+^2}{1+3f_+^2}\right)^{1/2}\Big(\langle\yop^2\rangle-\langle\zop^2\rangle\Big).$$ For $\lambda\geq \frac{3}{2\sqrt{2}}$, the second term is bounded and of order $1$, \textit{including at the critical point}. Therefore, since $\bar{X}\gg 1$, we can write $\bra{G_{+++}}\xop_a\xop_b\xop_c\ket{G_{+++}}=8\bar{X}^3+O(\bar{X})$; the dominant term in the expression will be the same as in the absence of quantum fluctuations. The same is true for all of the terms appearing in $\mathcal{C}_{abc}$; the correction due to the quantum fluctuation will always be sub-dominant. In the end, we will get $$\mathcal{C}_{abc}=\frac{8\bar{X}^3+O(\bar{X})}{8\lvert\bar{X}\rvert^3+O(\bar{X}^2)}\sim -1.$$ Therefore, the superposition of four displaced squeezed state cannot account for the divergence of the coskewness at the critical point, as the quantum fluctuations predicted by the Gaussian analysis are bounded. This shows that in proximity of the critical point the system develops genuine non-Gaussian features.\\

\section{Details on the numerical simulations}

\subsection{Hamiltonian numerical simulations}

To resolve the spectral features of the Hamiltonian model, we resort to exact diagonalization.
This means that we fix a cutoff $C$ such that, for any $m>C$ or  $n >C$ we set 
$\bra{m_j}\hat{H} \ket{n_j} =0$, where $\ket{m_j}$ and $\ket{n_j}$ represent Fock states with $m$ or $n$ photons in the $j$-th cavity.
To verify the convergence of our results, we compare the results obtained for a cutoff $C$ and those obtained for a cutoff $C' = C + \eta +2$.
In particular, we verify that the eigenvectors $\ket{\Psi_j(C)}$ associated with the lowest 10 eigenenergies $E_j$ and obtained with a cutoff $C$ are less than $0.5\%$ different with respect to those obtained with a cutoff $C'$, i.e.,
\begin{equation}
    |\bracket{\Psi_j(C)}{ \Psi_j(C')}|< 0.005.
\end{equation}

In the exact diagonalization algorithm, we exploit both the $\mathbb{Z}_2 \otimes \mathbb{Z}_2$ symmetry of $\hat{H}$, and the translational invariance of the Hamiltonian, to transform $\hat{H}$ into its block diagonal form. We then diagonalize each one of the blocks, obtaining the eigenvalues and eigenvectors associated with each one of these symmetry sectors.
Let us also notice that this procedure allow us determining the correct form of the ground and excited states even in regimes where numerical errors would make impossible to distinguish between them. Indeed, from analytical considerations, we know that, for any finite-size system, phase transitions cannot occur without thermodynamical or parameter rescaling limit. Therefore, the eigenvalues of the Hamiltonian can never become truly degenerate.

\subsection{Dissipative numerical simulations}

To investigate the dissipative model, we resort to quantum trajectories.
 A quantum trajectory (also known as wave function Montecarlo) is a mapping of the Lindblad master equation onto a stochastic differential equation for the wave function $\ket{\psi(t)}$. 
 The wave function $\ket{\psi(t)}$ has a piece-wise deterministic evolution under the action of a non-Hermitian Hamiltonian $\hat{H}_{\rm eff}$, which in our case reads 
\begin{equation}
     \hat{H}_{\rm eff} = \hat{H} - i \frac{\gamma}{2}
     \left( \hat{a}^\dagger \hat{a} + \hat{b}^\dagger \hat{b} + \hat{c}^\dagger \hat{c} \right),
\end{equation}
randomly interrupted by the occurrence of quantum jumps $\hat{J}$ (one of the three operators $\hat{a}$, $\hat{b}$, or $\hat{c}$).
In an infinitesimal time-step $dt$, each quantum jump occurs with a probability $p_J = \gamma dt \expec{\hat{J}^\dagger \hat{J}}$,
where $\hat{J} \in \{ \hat{a}, \, \hat{b}, \, \hat{c} \}$.

A numerical simulation of a quantum trajectory is thus equivalent to extract the probability that a quantum jump occurred at each time-step $dt$. 
If no quantum jump occurs, then
\begin{equation}
     \ket{\psi(t+dt)} = \ket{\psi(t)} 
     - i dt \hat{H}_{\rm eff} \ket{\psi(t)}.
 \end{equation}
Otherwise, according to the quantum jump extracted, the evolution is given by
\begin{equation}
\ket{\psi(t+dt)} 
     = \frac{\hat{J} \ket{\psi(t)}}{\expec{ \psi(t)| \hat{J}^\dagger \hat{J}| \psi(t)}}.
 \end{equation}
 The results of the Lindlbad master equation can be then retrieved by averaging over a large number $N_{\rm traj}$ of quantum trajectories, because
 \begin{equation}
     \rhot = \lim_{N_{\rm traj} \to \infty} \sum_{j=1}^{N_{\rm traj}} \frac{\ket{\Psi(t)}\bra{\Psi(t)}}{N_{\rm traj}}.
 \end{equation}

The advantage of quantum trajectory is thus to reduce the numerical cost of a single simulation (from that of a density matrix, to one of a wave function), but the price to pay is the need to perform the simulation several times.
To obtain the results shown in the main text, we exploited the parellalizable nature of quantum trajectories.
Let us notice that, for the largest $\eta$ considered here, i.e., $\eta=3$, we were able to obtain reliable results only when considering a cutoff $C = 25$ ($C'=30$), and thus a Hilbert space of dimension $15625$ ($27000$). This demonstrates the very non-Gaussian photon-number distribution of the states across the transition.
Furthermore, at the transition phenomena such as hysteresis require (i) long simulations and (ii) increase the mixed nature of the steady state, making it necessary to increase the number of quantum trajectories to reduce the statistical noise.

\color{black}

\end{document}